\documentclass[a4paper,fleqn,usenatbib,useAMS]{mnras}

\usepackage{graphicx}	% Including figure files
\usepackage{amsmath}	% Advanced maths commands
\usepackage{amssymb}	% Extra maths symbols
\usepackage{multicol}        % Multi-column entries in tables
\usepackage{bm}		% Bold maths symbols, including upright Greek
\usepackage{pdflscape}	% Landscape pages
\usepackage[encapsulated]{CJK}
\usepackage{ucs}
\usepackage[utf8x]{inputenc}
\usepackage[T1]{fontenc}
\usepackage{ae,aecompl}

\usepackage{newtxtext,newtxmath}
\title[Extended X-rays from DQ Her]{Extended X-ray emission from the classic nova DQ\,Her - On the possible presence of a magnetized jet}
\author[J.A.\,Toal\'{a} et al.]{J.A.\,Toal\'{a}$^{1}$\thanks{E-mail:\,j.toala@irya.unam.mx}, M.A.\,Guerrero$^{2}$, E.\,Santamar\'{i}a$^{3}$, G.\,Ramos-Larios$^{3}$ and L.\,Sabin$^{4}$\\
$^{1}$Instituto de Radioastronom\'{i}a y Astrof\'{i}sica (IRyA), UNAM Campus Morelia, Apartado postal 3-72, 58090 Morelia, Michoacan, Mexico\\
$^{2}$Instituto de Astrof\'{i}sica de Andaluc\'{i}a (IAA-CSIC), Glorieta de la Astronom\'{i}a S/N, 18008 Granada, Spain\\
$^{3}$Instituto de Astronom\'{i}a y Meteorolog\'{i}a, Universidad de Guadalajara, Av. Vallarta 2602, Arcos Vallarta, 44130 Guadalajara, Mexico\\
$^{4}$Instituto de Astronom\'{i}a,  Universidad Nacional Autónoma de México, Apdo. Postal 877, 22800 Ensenada, B.C., Mexico
}

\pubyear{2018}

\begin{document}
\label{firstpage}
\pagerange{\pageref{firstpage}--\pageref{lastpage}}
\maketitle

\begin{abstract}

We present an analysis of archival \emph{Chandra} and \emph{XMM-Newton} 
observations of the magnetically-active cataclysmic variable DQ\,Her and 
the shell around it ejected in a nova event in 1934.  
A careful revision of the \emph{Chandra} observations confirms previous 
claims on the presence of extended X-ray emission around DQ\,Her and 
reveals that it actually corresponds to a bipolar jet-like structure 
extending $\simeq$32$''$ along a direction from NE to SW.  
Therefore, this X-ray emission extends beyond the optical nova shell 
and is perpendicular to its major axis. 
The \emph{XMM-Newton} observations confirm the presence of the extended X-ray 
emission detected by \emph{Chandra}, suggesting the additional presence of a 
diffuse X-ray emission from a hot bubble filling the nova shell. 
This hot bubble was very likely produced by the explosion that created the 
nebular shell detected in optical images.
The bipolar feature can be modelled by the combination of an optically 
thin plasma emission component with temperature $T\approx2\times10^{6}$~K 
and a power law component with a photon index of $\Gamma=1.1\pm0.9$.  
Its X-ray luminosity in the 0.3--5~keV energy range is 
$L_\mathrm{X}=(2.1\pm1.3)\times10^{29}$~erg~s$^{-1}$, 
for an electron density $n_\mathrm{e}\approx2$~cm$^{-3}$ 
and a mass $m_\mathrm{X}\approx 3\times10^{-6}$~M$_{\odot}$. 
%consistent with the expectations of a nova explosion. 
We suggest that the X-ray bipolar structure in DQ\,Her is a jet and 
interpret its non-thermal X-ray emission in terms of a magnetized jet.

\end{abstract}


%%%%%%%%%%%%%%%%%%%%%%%%%%%%%%%%%%%%%%%%%%%%%%%%%%

%%%%%%%%%%%%%%%%% BODY OF PAPER %%%%%%%%%%%%%%%%%%

% The MNRAS class isn't designed to include a table of contents, but for this document one is useful.
% I therefore have to do some kludging to make it work without masses of blank space.
%\begingroup
%\let\clearpage\relax
%\tableofcontents
%\endgroup
%\newpage

\section{INTRODUCTION}
\label{sec:intro}

%Low-mass stars are the most numerous stars in the Universe. These are
%responsible for enriching the observed interstellar medium of their
%hosts galaxies through their stellar winds and explosions.

%In Santamar\'{i}a et al. (in prep.) we study the time-evolution of
%classic novae using multi-epoch optical observations of a sample of 6
%novae, including DQ\,Her. We have shown that even after a XX~yr of
%time, these objects expand linearly with radius, that is, in a
%free-expanding stage similarly to the early stage of evolution of a
%supernova remanant. In particular DQ\,Her has an expansion rate of
%XX~$''$~yr$^{-1}$.

The detection of extended X-ray emission from classical novae (CNe)
has proven to be rare. 
Thorough archival studies searching for diffuse X-ray emission from nova 
shells have been presented in the past \citep[e.g.,][]{Orio2001,Balman2006}, 
but there is only a handful number of novae with reported extended X-ray 
emission. 

The first detection of extended X-ray emission in a CN was obtained for 
GK\,Per using \emph{ROSAT} PSPC observations \citep{Balman1999}. Subsequent observations of GK\,Per by the \emph{Chandra} X-ray observatory 
\citep[][]{Balman2005} demonstrated that its extended X-ray 
emission can be described by a non-equilibrium thermal plasma 
component with additional synchrotron emission.  
With a total energy $\sim$10$^{-7}$ times that of a classic supernova 
explosions ($\sim$10$^{51}$~erg~s$^{-1}$), the diffuse X-ray emission 
from GK\,Per is a scaled down version of those events.  
More recent \emph{Chandra} observations showed that the X-ray brightness of 
GK\,Per declines with time as a result of its expansion \citep{Takei2015}, 
implying that the diffuse X-ray emission from CNe is short-lived.  
The dramatic morphological and spectral variations of its 
X-ray emission revealed by \emph{Sukaku} observations probe 
the interactions of this nova remnant through its complex 
circumstellar medium \citep{Yuasa2016}.

Extended X-ray emission has also been reported in \emph{Chandra} observations of 
RR\,Pic \citep{Balman2004}, although the marginal detection ($\sim$60~photons) 
makes difficult an assessment of the spatial correlation of the extended X-ray 
emission with the optical nova shell \citep[see figure~1 in][]{Balman2006}.  
% In spite of the small count number, the X-ray image suggests that the 
% X-ray-emitting gas is oriented along the north-south direction, i.e., 
% the same as the major axis of the elliptical nebula detected in 
% H$\alpha$ narrow-band images \citep[see figure~1 in][]{Balman2006}. 
Marginal detections of extended X-ray emission have been claimed for the 
recurrent nova T\,Pyx \citep{Balman2014} and the cataclysmic variable (CV) DK\,Lac \citep{TSD2013}, but
the former has been questioned \citep{MSN2012}.  
% Diffuse X-ray emission extending up to 10$''$ from the progenitor star of the 
% CN T\,Pyx has also been reported \citep{Balman2010} using \emph{XMM-Newton} 
% observations, although the spatial resolution of \emph{XMM-Newton} can only 
% marginally resolve the diffuse emission from that of the bright central 
% source. 
Finally, an extended 1\farcs2 jet-like feature in the soft (0.3--0.8~keV) 
energy band has been reported in \emph{Chandra} observations of the 
recurrent nova RS\,Oph \citep{Luna2009}.  
The orientation of this extended X-ray emission is consistent with the radio 
and IR emission from the ring of synchrotron-emitting plasma associated with 
the most recent blast wave \citep[][]{Chesneau2007}.
%, although some instrumental 
%effects casts doubts on the true nature of this feature.  

In this work we focus on the extended X-ray emission from DQ\,Her, 
a slow nova from a CV system that experienced 
an outburst on December 1934 and ejected a nova shell with a present 
angular size of 32$^{\prime\prime}\times$24$^{\prime\prime}$ 
\citep{Santamaria2020}. 
This classical nova was not initially detected by \emph{Einstein} 
\citep{Cordova1981}, but \citet{Silber1996} reported its detection 
in \emph{ROSAT} Position Sensitive Proportional Counters (PSPC) observations with an X-ray luminosity in the 0.1--2.0~keV energy range of $4\times10^{30}$~erg~s$^{-1}$. 
The low number of photons detected in these observations precluded at that 
time a detailed characterization of the X-ray properties from DQ\,Her. 
Higher quality \emph{Chandra} Advanced CCD Imaging Spectrometer\,(ACIS)-S observations were used by 
\citet{Mukai2003} to study the spectral properties and time 
variation of the X-ray-emitting, magnetically-active progenitor 
star of DQ\,Her \citep[e.g.,][]{Walker1956}. 
\citet{Mukai2003} found that the best-fit model to the X-ray spectrum of the 
progenitor star of DQ\,Her is composed by an optically-thin plasma emission 
model plus a power-law, the latter component in line with the magnetic field 
of DQ\,Her. 
Furthermore, their analysis of radial profiles of the X-ray emission hinted 
at the presence of extended X-ray emission with energies below 0.8~keV at 
distances up to $\sim10''$ from DQ\,Her that they associated with individual 
clumps in the nova shell.

We present here a joint analysis of archival \emph{XMM-Newton} European Photon Imaging Camera (EPIC)  
observations and revisit the \emph{Chandra} ACIS-S observations of 
DQ\,Her. 
The combination of both archival data confirms that the extended X-ray 
emission from DQ\,Her is indeed real and originates from emission 
filling the nova shell and a bipolar (jet-like) feature. 
This paper is organised as follows. 
In Section~2 we describe the observations analysed here.  
The results of the imaging and spectral analyses are presented in Section~3 
and 4, respectively. 
A discussion of our results is presented in Section~5 and 
a summary in Section~6.

\section{Observations and data preparation}

\subsection{{\it Chandra} Observations}

\begin{figure}
\begin{center}
  \includegraphics[angle=0,width=\linewidth]{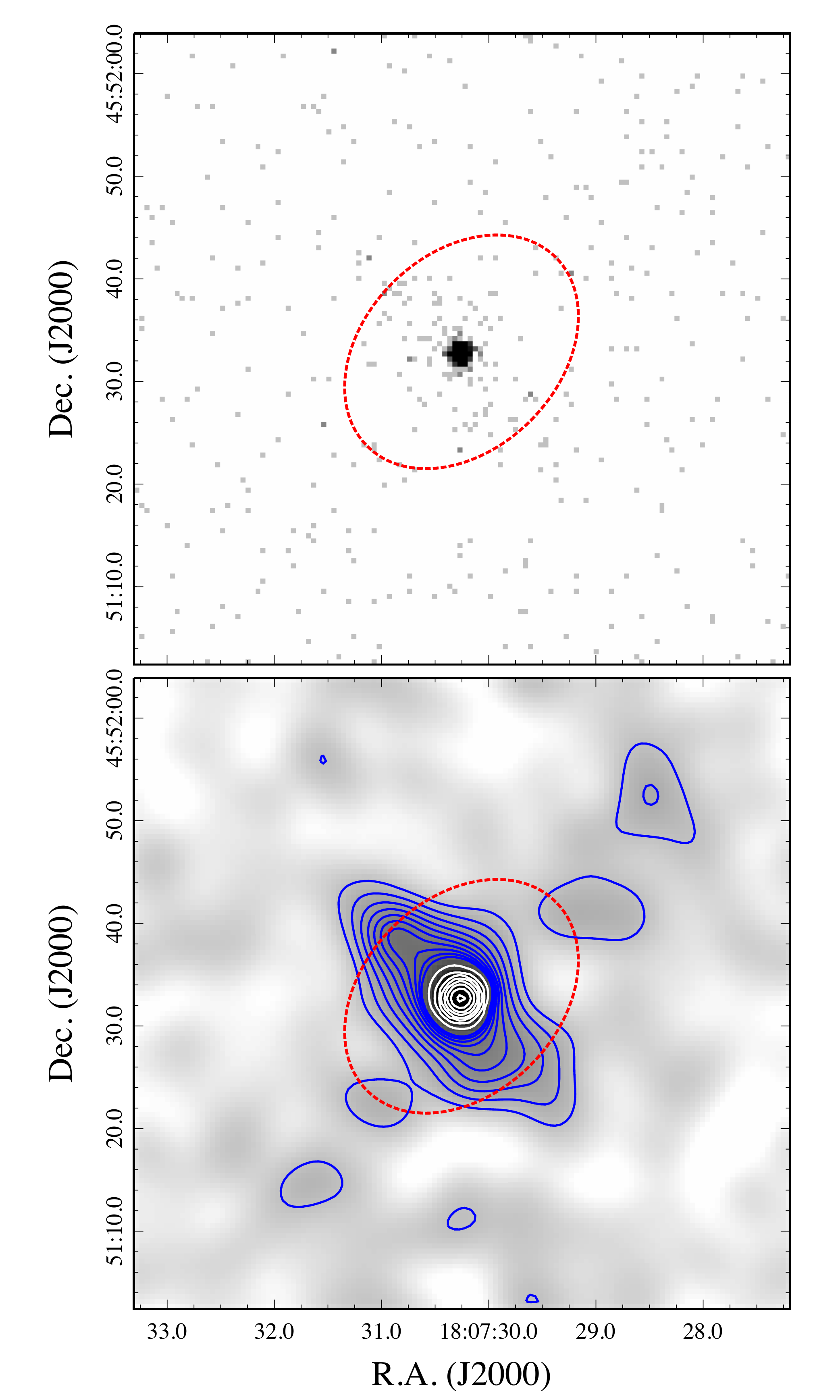}
\caption{\emph{Chandra} ACIS-S images of DQ\,Her. 
Top panel: An image of the event file using the natural 0\farcs5 in size 
ACIS-S pixel. Bottom: Adaptively smoothed image of the extended X-ray emission in DQ\,Her 
detected by \emph{Chandra}. Both images were obtained in the 0.3--5.0~keV energy range. The dashed ellipse shows the extent of the optical nebula as presented in Figure~3.}
\label{fig:DQ_Her_Xrays1}
\end{center}
\end{figure}

DQ\,Her was observed by the \emph{Chandra} X-ray Observatory with a
total exposure time of 70~ks split into two observations performed 
on 2001 July 26 and 29. 
The back-illuminated S3 CCD on the ACIS-S was used for these observations (Obs.\,ID.\ 1899 and 2503, PI: K.\,Mukai). 
The ACIS-S data were reprocessed with the \emph{Chandra} Interactive
Analysis of Observations ({\sc ciao}) software \citep[version
  4.11;][]{Fruscione2006}. % and are presented here for
%discussion and comparison with the \emph{XMM-Newton}
%observations.
After combining the data and excising high-background and dead periods 
of time, the net exposure time was 68~ks. 
X-ray images of DQ\,Her obtained after combining the two data sets 
are presented in Figure~\ref{fig:DQ_Her_Xrays1}.  

The processed event \emph{Chandra} image of DQ\,Her in the 0.3--5.0 keV energy range 
(Fig.~\ref{fig:DQ_Her_Xrays1} - top) undoubtedly shows that the central 
star is a point-source of X-ray emission. 
To unveil the true extension of the diffuse X-ray emission in DQ\,Her, 
we created a smoothed image in the 0.3--5.0~keV energy range using 
the {\sc ciao} task {\it csmooth}. The smoothing process was performed using a Gaussian kernel and a fast-Fourier transform (FFT) convolution method. Regions in the event file above $3\sigma$-confidence levels remained unsmoothed, preventing the emission from the central star to be highly smoothed. 
The resultant image is shown in the bottom panel of 
Figure~\ref{fig:DQ_Her_Xrays1}, where a highly elongated 
extended emission along the NE-SW direction is clearly 
shown. Alternatively, we used the suite {\sc MARX} 5.5.0 \citep{Davis_etal2012} to model the \emph{Chandra} point spread function (PSF) of a point-source with the spectral 
properties of the central star of DQ\,Her \citep[as described by][but see also 
section 4.1 below]{Mukai2003}. The comparison of this synthetic X-ray point-source with the image of DQ\,Her confirms the presence and extent of this diffuse emission.

\begin{figure*}
\begin{center}
  \includegraphics[angle=0,width=0.9\linewidth]{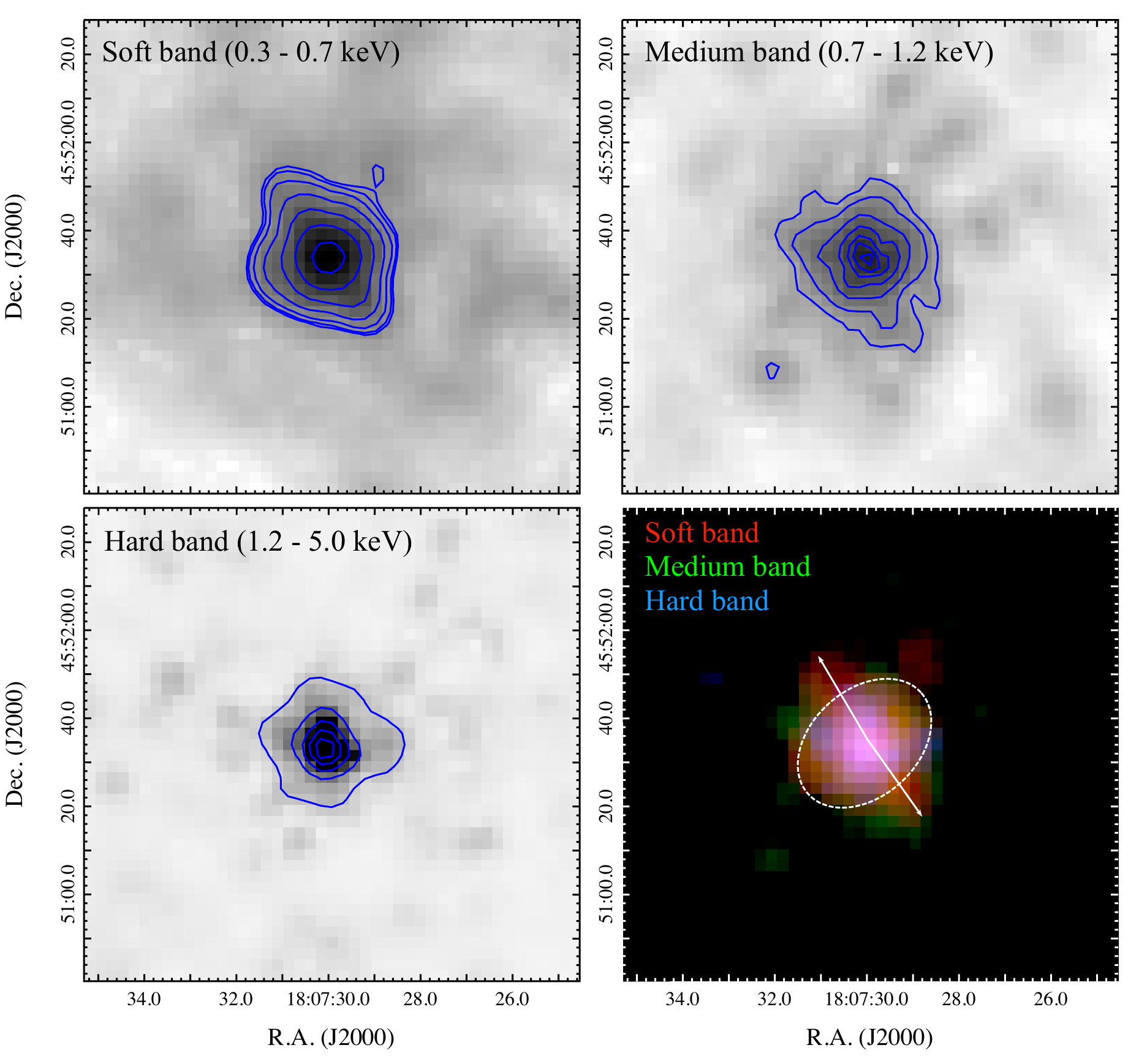}
\caption{\emph{XMM-Newton} EPIC (pn+MOS1+MOS2) images of DQ\,Her. The
  bottom right panel shows a colour-composite image obtained by
  combining the other three panels. Red, green and blue correspond to
  the soft, medium and hard bands. The extension of the nebular remnant
  detected in optical observations (see Fig.~3) is shown with an
  elliptical dashed-line region. The suggested bipolar structure is
  shown with arrows.}
\label{fig:DQ_Her_Xrays2}
\end{center}
\end{figure*}

The \emph{Chandra} spectra of the central star and that of the extended
emission were extracted separately from each ACIS observations using
the {\sc ciao} task {\it specextract}, which produces the
corresponding calibration matrices. 
The spectra and calibrations matrices from different data sets were 
subsequently merged using the {\sc ciao} task {\it combine\_spectra}. 
The spectrum of the central star of DQ\,Her was extracted from a 
circular aperture with radius of 2$''$ and that of the extended 
X-ray emission from an elliptical aperture with semi-minor and 
major axes 10$''$ and 18$''$ encompassing the emission detected 
in the bottom panel of Figure~\ref{fig:DQ_Her_Xrays1}.  
The emission from the central source was excised from the latter.     
% To avoid any possible contribution from the extended X-ray emission,  
% the background was extracted from an 
% annular region with inner and outer radii around the star 
% of 4.5$''$ and 10$''$, respectively. 
% The emission from the central star was excised from a circular 
% aperture around it, and t
The background was extracted from a region without contribution from 
extended emission nor point sources. 
The net count rates in the 0.3--5.0~keV energy range 
are 22.6~counts~ks$^{-1}$ for the central star 
and 1.32~counts~ks$^{-1}$ for the extended emission 
for total count numbers $\simeq$1,500 and $\simeq$90 
counts, respectively.
% 22.6 x 68 = 1536 cnts
% 1.32 x 68 =   90 cnts 

\subsection{{\it XMM-Newton} Observations}

DQ\,Her was observed by \emph{XMM-Newton} on 2017 April 19 with the
three EPIC cameras for a total exposure time of
41.9~ks (PI: H.\,Worpel; Obs.\,ID.: 0804111201). 
The EPIC pn, MOS1 and MOS2 cameras were operated in the Full Frame Mode with 
the thin optical blocking filter. 
The individual observing times for the pn, MOS1, MOS2, and pn 
cameras were 39.0~ks, 40.6~ks, and 40.5~ks, respectively.
The \emph{XMM-Newton} data were processed with the Science Analysis
Software ({\sc sas}; version 17.0), using the {\it epproc} and 
{\it emproc} SAS tasks to apply the most recent calibrations 
available on February 2020.  
After excising periods of high background, the total useful time of the 
pn, MOS1 and MOS2 cameras were 15.2~ks, 25.4~ks and 26.2~ks, respectively.

We used the Extended Source Analysis Software ({\sc esas}) tasks 
to map the distribution of the X-ray-emitting gas in DQ\,Her. Background-subtracted, exposure-corrected EPIC pn, MOS1, and MOS2 
images were created and merged. 
EPIC images in the soft 0.3--0.7~keV, medium 0.7--1.2~keV, and
hard 1.2--5.0~keV energy bands were created.  
The individual images and a colour-composite X-ray picture are 
presented in Figure~\ref{fig:DQ_Her_Xrays2}.

Spectra and their corresponding associated callibration matrices were 
obtained from a circular aperture with radius of 24$''$ centered on the 
central star of DQ\,Her using the {\it evselect}, {\it arfgen} and 
{\it rmfgen} {\sc sas} tasks. 
Due to the lower spatial resolution of the EPIC cameras compared 
to that of ACIS-S, the contribution from the central star cannot 
be properly resolved from that of the extended X-ray emission.  
Therefore, the EPIC spectra encompass the emission from 
both the point-source and extended component of DQ\,Her. 
The net count rates of the pn, MOS1 and MOS1 cameras are 41.9~counts~ks$^{-1}$, 
8.4~counts~ks$^{-1}$, and 11.6~counts~ks$^{-1}$, respectively.

\begin{figure*}
\begin{center}
  \includegraphics[angle=0,width=0.75\linewidth]{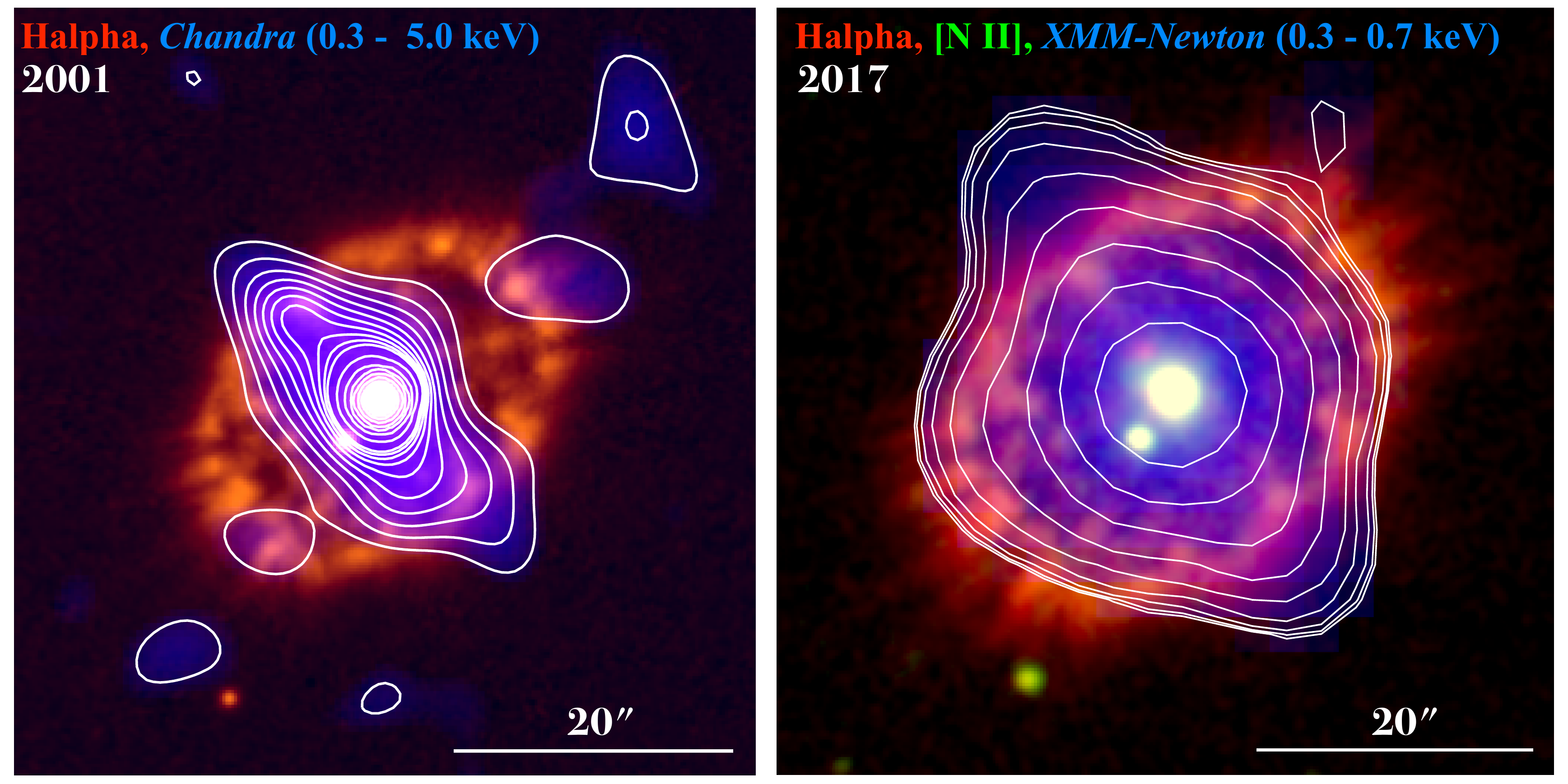}
\caption{Comparison between the optical emission from DQ\,Her and the X-ray observations obtained with \emph{Chandra} (left) and \emph{XMM-Newton} (right). The nebular images from 2017 were obtained at the NOT and the corresponding H$\alpha$ 2001 image was obtained by expanding a WHT from 1997 by 6\% according to \citet{Santamaria2020} (see Section~3 for details).}
\label{fig:DQ_Her_Xrays3}
\end{center}
\end{figure*}

%\subsection{Optical images}

\section{Extended X-ray emission from DQ\,Her}

The \emph{Chandra} ACIS-S images presented in Figure~\ref{fig:DQ_Her_Xrays1} 
clearly confirms the presence of extended emission as previously reported by 
\citet{Mukai2003}. 
% This emission is detected beyond 10$''$ from DQ\,Her.  
More importantly, the X-ray image in the bottom panel of this figure shows 
that this emission has a bipolar, jet-like shape $\simeq$16$''$ wide and 
$\simeq$32$''$ long in the NE to SW direction (PA$\approx$45$^\circ$). 
The direction of this bipolar feature is orthogonal to the apparent 
semi-major axis of the nebula. The \emph{Chandra} X-ray image shows that the bipolar feature presents hints of an S-shape, more clearly seen in its SW section.

The \emph{XMM-Newton} EPIC images presented in Figure~\ref{fig:DQ_Her_Xrays2} 
also show extended X-ray emission. 
The images in the three X-ray bands present a peak at the location 
of the central star of DQ\,Her, as well as extended emission that, 
depending on the X-ray band, uncover different features.
The soft (0.3--0.7~keV) EPIC X-ray image 
(Fig.~\ref{fig:DQ_Her_Xrays2} top left) 
is indicative of
a bipolar morphology protruding from the central star and
extending towards the NE and SW directions, very similar to that 
of the extended X-ray emission detected in the \emph{Chandra} 
images.  
The diffuse emission detected in the medium (0.7--1.2~keV)
EPIC X-ray band (Fig.~\ref{fig:DQ_Her_Xrays2} - top right) 
also extends towards the NE and SW regions, but this component 
is not spatially coincident with the bipolar features detected 
in the soft band. 
Instead, it seems to surround the soft emission. 
Finally, the spatial distribution of the emission in the hard (1.2--5.0~keV) 
EPIC X-ray band is more centrally-concentrated than in the other two EPIC
bands (see Fig.~\ref{fig:DQ_Her_Xrays2} - bottom left) and is basically 
consistent with a point-source with some contribution to the extended emission.  
All these characteristics are illustrated in the colour-composite X-ray 
picture presented in the bottom right panel of Figure~\ref{fig:DQ_Her_Xrays2}.  

To further peer into the spatial distribution of the X-ray-emitting material
in DQ\,Her, we show in Figure~\ref{fig:DQ_Her_Xrays3} a comparison between 
the X-ray and narrowband optical images. 
As noted by \citet{Santamaria2020}, DQ\,Her has an angular expansion rate 
sufficiently large (0\farcs188 yr$^{-1}$ along its major axis) to result 
in a noticeable angular expansion within a few years. 
To produce consistent comparisons between the optical nebular remnant 
and the X-ray images, optical images obtained at similar epochs than 
those of the X-ray images have to be used.  
For comparison with the \emph{Chandra} images, the closest contemporary
available image of DQ\,Her is an H$\alpha$ image taken at the William 
Herschel Telescope (WHT, La Palma, Spain) on 1997 October 25, i.e., 
about 4 years before the X-ray observation.  
An expansion factor of 6\% was applied to this optical image, following the 
expansion rate reported by \citet{Santamaria2020}, to produce a synthetic 
2001 H$\alpha$ image suitable for comparison with the \emph{Chandra} X-ray 
image (Fig.~\ref{fig:DQ_Her_Xrays3} - left). 
For comparison with the \emph{XMM-Newton} images, we used H$\alpha$ and 
[N\,{\sc ii}] images obtained at the Nordic Optical Telescope (NOT, La 
Palma, Spain) on 2017 May 27, just about one month after the 
X-ray observation (Fig.~\ref{fig:DQ_Her_Xrays3} - right).  
The inspection of the pictures in Figure~\ref{fig:DQ_Her_Xrays3} clearly reveals that the 
bipolar jet-like feature extends beyond the optical nebula 
and is oriented along its minor axis.

\section{Spectral Analysis}

\begin{figure*}
\begin{center}
\includegraphics[angle=0,width=0.8\linewidth]{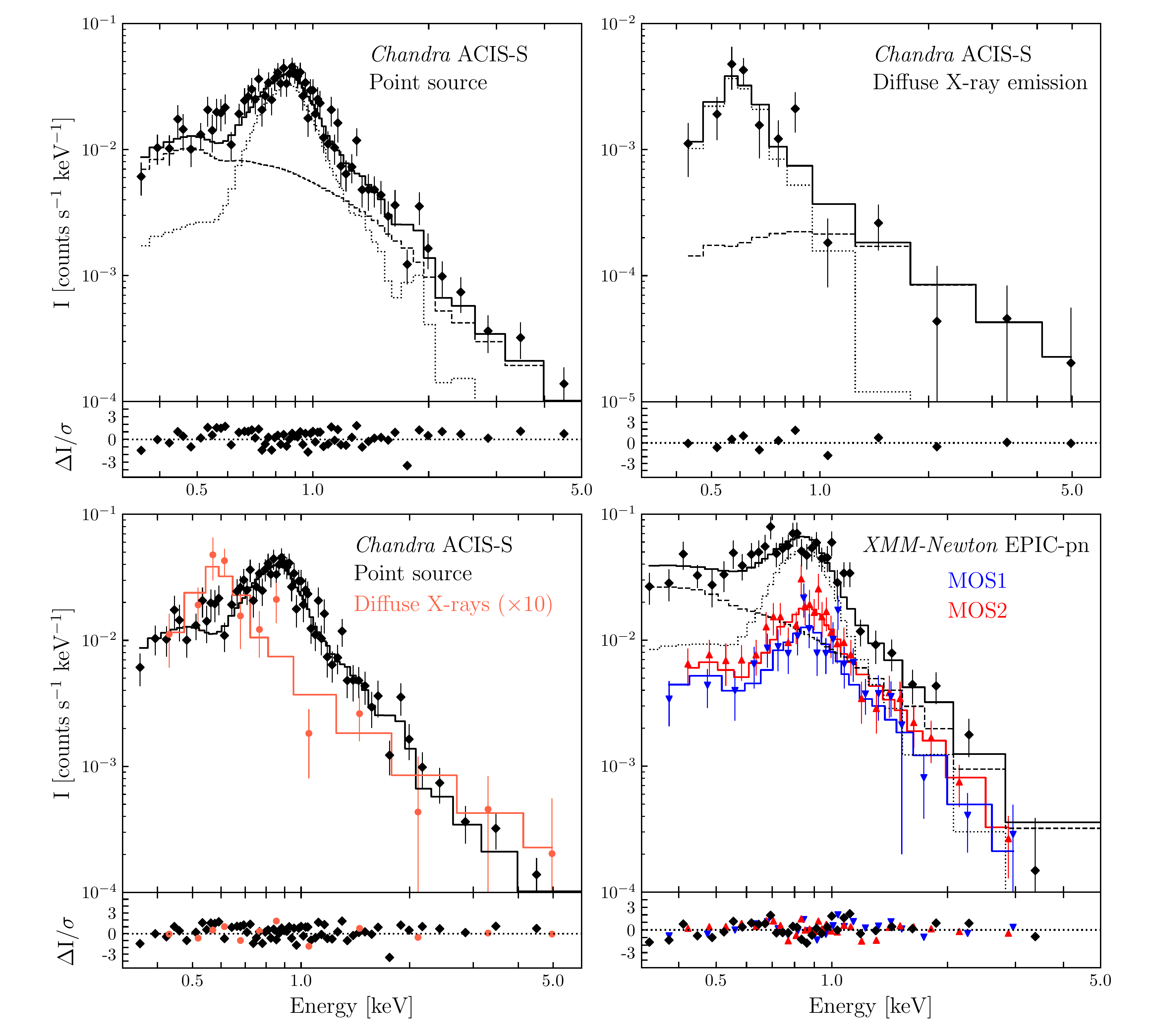}
\caption{Background-subtracted spectra of DQ\,Her. 
The top panels show the \emph{Chandra} ACIS spectra of the point-source 
(left) and diffuse emission (right).  
The two spectra are shown together in the bottom-left panel, where 
the emission from the diffuse component has been scaled for an easy 
comparison.  
The bottom-right panel presents the \emph{XMM-Newton} EPIC spectra 
of DQ\,Her, with different symbols and colours for the spectra extracted 
from different EPIC cameras. 
In all panels the solid lines represent the best-fit model to the data, 
while the dotted and dashed lines show the contribution of {\it apec} 
and non-thermal power-law components, respectively.}
\label{fig:DQ_Her_spec}
\end{center}
\end{figure*}

The superb angular resolution of the \emph{Chandra} ACIS camera has allowed us 
to extract individual spectra for the point source and extended X-ray emission 
from DQ\,Her (Fig.~\ref{fig:DQ_Her_spec}-top left and right, respectively). 
These ACIS spectra reveal remarkable spectral differences between the 
central and the diffuse emission, as illustrated in the bottom-left 
panel of Figure~\ref{fig:DQ_Her_spec}, regardless of the lower quality 
of the spectrum of the diffuse component.
As shown by \citet{Mukai2003}, the spectrum of the star peaks between 
0.8--1.0~keV, with some contribution to the soft energy range below 
0.7~keV. The spectrum then declines for energies above 1.0~keV showing the 
contribution from some spectral line very likely the Si\,{\sc xiii} 
at 1.8~keV. On the other hand, the spectrum of the extended emission peaks at softer 
energies, $\lesssim$0.6~keV, declining towards higher energies, with 
hints of the presence of spectral lines at 0.9~keV and 1.4~keV.  
The former could be attributed to the O\,{\sc vii} triplet at 0.58~keV, the 
Fe complex at $\approx$0.8~KeV, and the Ne\,{\sc ix} lines at 0.9 keV, and 
the latter to Mg\,{\sc xi} at 1.4~keV.

In the following we present the details of the spectral analysis of the {\it Chandra} ACIS-S and {\it XMM-Newton} EPIC spectra of DQ\,Her. The parameters of the best-fit models for the different spectra as well as their significance are listed in Figure~\ref{fig:DQ_Her_spec} are listed in Table~1.

%The maximum detected in the EPIC images is coincident with the
%position of the central star of DQ\,Her, but due to the large PSF of
%these cameras it is almost impossible to extract a spectrum that does
%not includes the contribution from both the central star and the
%extended X-ray emission.
%Thus, to study the physical properties of the X-ray emission from
%DQ\,Her we start by fitting the spectrum extracted from the {\it
%  Chandra} observations. The background-subtracted spectra of the
%central star and that of the exteded X-ray emission (see Fig.~1) are
%presented in Figure~3 left and middle panels.

\subsection{X-rays from DQ\,Her}

Following \citet{Mukai2003} we fitted a two-component model to the
\emph{Chandra} ACIS-S spectra of the central star of DQ\,Her 
consisting of an optically-thin {\it apec} emission model and a  
power-law component. The former component can be attributed to a hot plasma and 
the latter to non-thermal synchrotron emission. 
Our best fit model, which is presented in the left panel of 
Figure~\ref{fig:DQ_Her_spec} in comparison with the ACIS-S 
spectrum, has parameters consistent with those obtained by 
\citet{Mukai2003} (see Table~1). 
The non-thermal emission with a power-law index of $\Gamma=2.45$
dominates in the soft energy range below 0.7~keV and above 1.0~keV, with a 
contribution $\simeq$43\% to the total intrinsic flux.
% ($\chi^{2}$/DoF=1.12), consistent with the spectral model 
% presented in \citet{Mukai2003}, resulted in a column density of
% $N_\mathrm{H}=(3.4\pm2.0)\times10^{20}$~cm$^{-2}$, plasma temperature
% $kT=0.77\pm0.05$ ($\approx9\times10^{6}$~K), and photon index
% $\Gamma=2.45^{+0.50}_{-0.40}$. 
% The corresponding normalization parameters\footnote{
%  The normalization parameter is defined as $A \approx 10^{-14}\int n_\mathrm{e}^{2} dV/4 \pi d^2$, where $n_\mathrm{e}$ and $d$ are the electron number density and the distance, respectively.} 
% are $1.33\times10^{-5}$~cm$^{-5}$ for the {\it apec} model and 
% $1.05\times10^{-5}$~cm$^{-5}$ for the power-law component. 
% The observed flux in the 0.3--5.0~keV energy range is
% $f_\mathrm{X}=(8.0\pm1.1)\times10^{-14}$~erg~s$^{-1}$~cm$^{-2}$, 
% which corresponds to an 
The intrinsic flux of DQ\,Her is found to be 
$F_\mathrm{X}=(9.4\pm1.2)\times10^{-14}$~erg~s$^{-1}$~cm$^{-2}$, 
implying a luminosity $L_\mathrm{X}=(2.8\pm0.4)\times10^{30}$~erg~s$^{-1}$ 
at a distance of 501$\pm$6~pc \citep{Schaefer2018}.

\subsection{Extended X-ray emission}

The spectrum of the extended bipolar X-ray emission detected in the 
\emph{Chandra} ACIS-S observations was initially fitted by a single 
optically thin {\it apec} emission model, but resulted in a poor 
quality fit with $\chi^{2}$/DoF=1.60. 
A power-law model results in an even worse fit ($\chi^{2}/$DoF$>$2.4), 
whereas a two-temperature plasma emission model does not improve the 
fit ($\chi^{2}$/DoF=1.60), as it is not able to appropriately fit the 
spectrum for energies above 2~keV. 
The best-fit model is achieved by using a similar model as that for the 
central star, that is, a plasma emission model plus a non-thermal power-law.  
This best-fit, whose parameters are listed in Table~1, is shown in 
the middle panel of Figure~\ref{fig:DQ_Her_spec} in comparison with 
the ACIS-S spectrum. 
% The best-fit ($\chi^{2}$/DoF=1.16) resulted in a hydrogen column density 
% of $N_\mathrm{H}=(1.8\pm0.2)\times10^{20}$~cm$^{-2}$ with plasma temperature 
% of $kT=0.18^{+0.05}_{-0.07}$~keV (that is, $T\approx2\times10^{6}$~K) and 
% photon index of $\Gamma=1.1\pm0.9$. 
% The normalization parameters are $2.3\times10^{-6}$~cm$^{-5}$ and
% $3.9\times10^{-7}$~cm$^{-5}$ for the {\it apec} and power-law
% component, respectively. 
% The corresponding absorbed and unabsorbed fluxes in the 0.3--5.0~keV energy 
% range are $f_\mathrm{X,diff}=(6.1\pm4.1)\times10^{-15}$~erg~s$^{-1}$~cm$^{-2}$
% and $F_\mathrm{X,diff}=(7.0\pm4.7)\times10^{-15}$~erg~s$^{-1}$~cm$^{-2}$,
% respectively. 
This panel shows that the {\it apec} component dominates the emission for 
energies below 1.0~keV and the power-law component for greater energies.
Indeed, the non-thermal component contributes to 40\% of the total unabsorbed 
flux for the 0.3--5.0~keV energy range, but the optically-thin plasma 
component contributes to 75\% of the flux for energies between 0.3 and 1~keV.
The luminosity of the extended bipolar emission detected in \emph{Chandra} 
is $L_\mathrm{X,diff}=(2.1\pm1.3)\times10^{29}$~erg~s$^{-1}$.

\begin{table*}
\begin{center}
\caption{\label{tab:elines} Details of the spectral modelling of the X-ray observations of DQ\,Her.}
\setlength{\tabcolsep}{0.7\tabcolsep}
\begin{tabular}{llccccccccc}
\hline
        & Instrument & $N_\mathrm{H}$        & $kT$          & $A_{1}^\mathrm{b}$              & $\Gamma$  & $A_{2}^\mathrm{b}$              & $f_\mathrm{X}^\mathrm{c}$ & $F_\mathrm{X}^\mathrm{c}$ & $F_{2}/F_\mathrm{X}^\mathrm{d}$ &$\chi^2$/DoF\\
        &            & (10$^{20}$~cm$^{-2}$) &(keV)          & ($10^{-5}$~cm$^{-5}$)&           & ($10^{-5}$~cm$^{-5}$)& (cgs)          & (cgs)  &  &  \\
\hline
DQ\,Her & ACIS-S  & 3.4$\pm$2.0   & 0.77$^{+0.05}_{-0.05}$& 1.33    & 2.45$^{+0.50}_{-0.40}$ & 1.05 & 8.00$\pm$1.10 & 9.40$\pm$1.20 & 0.43 & 1.12\\
Extended& ACIS-S  & 1.8$\pm$0.2   & 0.18$^{+0.05}_{-0.07}$& 0.26    & 1.10$^{+0.09}_{-0.09}$ & 0.04 & 0.64$\pm$0.41 & 0.70$\pm$0.47 & 0.40 & 1.16\\
DQ\,Her$+$Extended & EPIC-pn & 3.4 & 0.78$^{+0.06}_{-0.07}$& 1.15    & 2.34$^{+0.25}_{-0.21}$ & 1.20 & 7.60$\pm$1.10 & 8.90$\pm$1.00 & 0.60 & 1.08\\
DQ\,Her$+$Extended & EPIC(pn+MOS)$^\mathrm{a}$& 3.4 & 0.77$^{+0.04}_{-0.05}$& 1.20& 2.37$^{+0.20}_{-0.20}$ & 1.20 & 7.60$\pm$1.10 & 8.90$\pm$0.90 & 0.60 & 0.90\\
\hline
\end{tabular}\\
$^\mathrm{a}$Joint model fit of the EPIC pn, MOS1 and MOS2 spectra.\\
$^\mathrm{b}$The normalization parameter is defined as $A \approx 10^{-14}\int n_\mathrm{e}^{2} dV/4 \pi d^2$, where $n_\mathrm{e}$ and $d$ are the electron number density and the distance, respectively.\\
$^\mathrm{c}$ The fluxes are computed for the 0.3--5~keV energy range and are presented in $10^{-14}$~erg~s$^{-1}$~cm$^{-2}$ units.\\
$^\mathrm{d}$The $F_{2}/F_\mathrm{X}$ ratio represents the contribution from the power-law component to the total flux.
\end{center}
\end{table*}

To estimate an electron density for the extended X-ray emission we
used the definition of the normalization parameter (see Table~1) adopting an
ellipsoidal morphology with semi-axes of 8$''$, 8$''$ and 16$''$. The
electron density of the bipolar emission is estimated to be 
$n_\mathrm{e}\approx2$~cm$^{-3}$,
which corresponds to a mass of the X-ray-emitting material 
$\approx3\times$10$^{-6}$~M$_\odot$, well below the typical 
mass ejecta of nova events \citep[e.g.,][]{Gehrz1998,DellaValle2020} and 
the ionised mass of DQ\,Her
\citep[$2.3\times$10$^{-4}$~M$_\odot$;][]{Santamaria2020}. 

% In line with what is found with material ejected from such events \citep[see][]{Soko2006}. \\

\subsection{The \emph{XMM-Newton} Spectra}

Due to the large PSF of the EPIC cameras it is not possible to extract 
independent spectra for the central star and extended X-ray emission of 
DQ\,Her. The three EPIC spectra are shown in the right panel of Figure~\ref{fig:DQ_Her_spec}. 
Similarly to the \emph{Chandra} ACIS spectrum, the \emph{XMM-Newton} EPIC 
spectra of DQ\,Her show a main peak for energies around 0.8--1.0~keV with 
a secondary contribution for energies below 0.7~keV.  
In addition, these spectra show some contribution from the N\,{\sc vi} 
triplet at 0.43~keV that is not detected in the \emph{Chandra} spectrum 
due to its lower sensitivity at softer energies.  

We first modelled the EPIC-pn spectrum of DQ\,Her because of its larger 
count rate than that of the MOS cameras (see Section~2). 
For simplicity, we fixed the column density value to
that obtained to the best-fit model to the ACIS-S data
($N_\mathrm{H}=3.4\times10^{20}$~cm$^{-2}$) and adopted 
a similar model consisting of an {\it apec} plasma 
emission model for hot gas and a power-law component for non-thermal emission.
The best-fit model is consistent with that obtained for the \emph{Chandra} 
ACIS-S spectrum of the central star of DQ\,Her (see Table~1).  
% The best-fit model ($\chi^{2}$/DoF=1.08), which is very similar to that 
% obtained for the \emph{Chandra} ACIS-S spectrum of the central star of 
% DQ\,Her, accounts for an {\it apec} plasma emission and a power-law with 
% temperature $kT=0.78^{+0.06}_{-0.07}$keV and photon index 
% $\Gamma=2.34^{+0.25}_{-0.21}$, respectively.  
% Their corresponding normalization parameters are 
% $1.15\times10^{-5}$ and $1.20\times10^{-5}$ for 
% the {\it apec} and power-law components, respectively. 
% The total absorbed and unabsorbed fluxes in the 0.3--5.0~keV energy range 
% are
% $f_\mathrm{X}=(7.6\pm1.0)\times10^{-14}$~erg~s$^{-1}$~cm$^{-2}$ and
% $F_\mathrm{X}=(8.9\pm1.0)\times10^{-14}$~erg~s$^{-1}$~cm$^{-2}$,
% respectively. 
The total luminosity in this model is
$L_\mathrm{X}=(2.7\pm0.3)\times10^{30}$~erg~s$^{-1}$, with the 
power-law component contributing 60\% to the total intrinsic flux.
Models simultaneously fitting the three EPIC spectra resulted in very 
similar best-fit parameters (Table~1).

% Finally, a last model was attempted by simultaneously fitting the
% three EPIC spectra. This model resulted in very similar model to that
% of the EPIC-pn spectra. The best-fit model ($\chi^{2}$/Dof=0.90)
% resulted in plasma temperature of $kT=0.77^{+0.04}_{-0.05}$~keV with a
% photon index of $\Gamma=2.37\pm0.20$. Their corresponding
% normalization parameters are $1.14\times10^{-5}$ and
% $1.20\times10^{-5}$ for the {\it apec} and power-law,
% respectively. 
% The intrinsic flux is
% $F_\mathrm{X}=(8.9\pm0.9)\times10^{-14}$~erg~s$^{-1}$~cm$^{-2}$, 
% which corresponds to an X-ray luminosity of
% $L_\mathrm{X}=(2.7\pm0.3)\times10^{30}$~erg~s$^{-1}$. 
% Similarly to the fit to the EPIC-pn spectrum, the power-law component 
% contributes to 60\% of the total luminosity.
% 
% 
% Table~1 presents a summary of the results for the spectral fits 
% performed to the X-ray data of DQ\,Her.

\section{Discussion}

The spatial and spectral analyses of the \emph{Chandra} and \emph{XMM-Newton} 
observations of DQ\,Her presented in Sections~3 and 4 reveal the presence of diffuse 
X-ray emission with spectral properties differing from those of the central 
star.  
Owing to its better angular resolution, $\sim$1$''$ at $\lesssim$1~keV,  \emph{Chandra} resolves more clearly the morphology of this emission, 
which is found to be elongated, extending $\approx$32$''$ along the 
NE-SW direction at PA$\approx45^\circ$ (Fig~\ref{fig:DQ_Her_Xrays1} 
and Fig.~\ref{fig:DQ_Her_Xrays3} - left) with a subtle S-shape.  
Contrary to \citet{Mukai2003}'s interpretation,  this emission is 
not associated with any specific clump in the nebular remnant, but 
it extends beyond the optical nova shell along its minor axis.

This morphology is confirmed in the \emph{XMM-Newton} images presented 
in Figures~\ref{fig:DQ_Her_Xrays2} and \ref{fig:DQ_Her_Xrays3} right panel, 
although at a coarse angular resolution ($\sim6''$ at $\lesssim$1~keV).   
The misalignment of the arrows marking the tips of the elongated 
structure in Figure~\ref{fig:DQ_Her_Xrays3} bottom right panel is indeed 
consistent with its S-shape in the \emph{Chandra} images.  
In addition, the \emph{XMM-Newton} EPIC image in the soft band is suggestive 
of extended X-ray emission filling the nova shell around DQ\,Her. 

The extended emission in the soft \emph{XMM-Newton} image can be 
attributed to thermal emission from hot plasma produced by the 
nova explosion, an adiabatically-shocked hot bubble homologous 
to supernova explosions.  
%The presence of this hot gas is also detected in the \emph{GALEX} 
%far-UV image of DQ\,Her\footnote{
%The {\it GALEX} data have been obtained from the Mikulski Archive for Space 
%Telescopes (MAST) at \url{https://archive.stsci.edu/} and corresponds to the Obs.\,ID. 6371337969029090040.} 
%presented in the right panel of Figure~\ref{fig:DQ_Her_Xrays3}. 
This hot bubble would be spatially coincident with the nebular 
remnant with no contribution to the bipolar structure detected in X-rays.
On the other hand, the origin of the bipolar structure is intriguing.  
It cannot be attributed to hot gas escaping the nova shell, because its 
non-thermal component is not expected from shock-heated plasma inside a 
hot bubble and because it projects along the minor axis of the nova shell 
whilst the shell is disrupted along its major axis at PA=$-45^\circ$ 
\citep{Vaytet2007}.

The production of the bipolar X-ray emission in DQ\,Her could be argued 
to be at the origin of the nova explosion.  
Three-dimensional numerical simulations tailored to similar events, such as 
outbursts in symbiotic stars \citep[see, e.g.,][]{Walder2008,Orlando2017}, 
might help interpreting the bipolar X-ray feature in DQ\,Her. 
In particular, the simulations presented in \citet{Orlando2017} to 
model the X-ray emission from the symbiotic star V745\,Sco only 17 
days after its outburst are able to produce bipolar ejections of 
X-ray-emitting gas. 
This X-ray emission arises from a non-isotropic blast wave 
produced instantaneously at the nova event.  
Its emission would then be thermal, with physical conditions at early times 
similar to those of the thermal component of the jet-like feature of DQ\,Her. 
However, the non-thermal emission and continuous collimation 
of the jet in DQ\,Her, $\sim$80~yr after the nova event 
\citep[][]{Santamaria2020}, make these models unsuitable for
the case of DQ\,Her.

It is interesting to note that the CV at the center of DQ\,Her 
belongs to the class of magnetically active intermediate polars 
(IP) exhibiting strong magnetic fields of the order of 1--10~MG 
\citep[see][and references therein]{Barrett2017}, which results in the
presence of a truncated accretion disk. 
Indeed \citet{Mukai2003} suggested that the X-ray emission from DQ\,Her is 
produced by scattered X-ray photons due to the presence of an accretion disk 
wind making it an unusual IP system. 
Furthermore, spectral mapping of DQ\,Her has revealed that the material 
in the disk is spiraling-in \citep{Saito2010}. 
Thus, we suggest that the elongated structure of DQ\,Her could be interpreted 
as a magnetized jet produced by {\it hoop stress} at the inner regions of the 
accretion disk as it is {\it threaded} by the vertical magnetic field 
\citep{Livio1997}. 
In this scenario, which has been proven feasible in stellar systems, 
such as the case of the protostellar object HH80 \citep{CG2010}, the 
jet would be continuously fed by material falling into the accretion 
disk and then ejected by the hoop stress.  
The non-thermal X-ray emission from the bipolar feature in DQ\,Her 
seems to support this scenario. Non-thermal radio emission would lend additional support, 
as typically found in jets of symbiotic stars \citep[e.g., CH\,Cyg;][]{Ketal2010}.  
However, an inspection of Jansky Very Large Array (JVLA) observations of DQ\,Her 
discloses the lack of radio emission from either the central 
source or an extended component \citep[see][]{Barrett2017}. 
We note that the synchrotron radiation from relativistic electrons close to 
the accreting white dwarf in CVs \citep{Chan1982} has been found to be highly 
variable, as it is the case of the central source of DQ\,Her 
\citep{Pavelin1994}.

The disk$+$jet phenomenon is found in a variety of astrophysical systems, from 
protostellar and young stellar objects \citep{CG2010,CG2012}, evolved low-mass 
stars \citep[][]{Sahai1998}, and massive X-ray binaries \citep[][]{vanK1992} to 
AGNs \citep[][]{AlonsoHerrero2018}. 
White dwarfs can act as the compact object for jet collimation, and 
indeed disk+jet systems have been found in symbiotic stars in which 
a white dwarf accretes material from a main sequence or red giant 
companion, for example, the well-studied R\,Aqr 
\citep{Ramstedt2018,Schmid2018,Melni2018} or MCW\,560 \citep{SS2009}.  
On the other hand, the conspicuous absence of jets in CVs has been explained 
in terms of the particular physical conditions in these systems \citep{SL2004}, 
although recent observational and theoretical results have found some evidence 
for transient jets \citep{CK2020}.

As for nova shells, \citet{Shara2012} suggested that an elongated 
structure towards the NE region of GK\,Per was a jet, but 
\citet{Harvey2016} demonstrated that it is not dynamically 
related and it has a low velocity. This leaves us only with the claims of jet-like structures in RS\,Oph and M31N 2008-12a, two recurrent novae. 
The presence of a jet in RS\,Oph is suggested by a jet-like morphological 
feature with an extent $\sim1''$ discovered in \emph{Chandra} X-ray 
observations \citep{Luna2009}. Meanwhile, the presence of a jet in M31N 2008-12a is supported by high-velocity $\sim$4600$\pm$600~km~s$^{-1}$ features detected in \emph{Hubble Space Telescope} 
(\emph{HST}) Space Telescope Imaging Spectrograph (STIS) spectra \citep{Darnley2017}, 
which are interpreted as an ejecta expanding in the direction close to the line of 
sight \citep[see also][and references therein]{Darnley2016}. 
Since the jet in RS\,Oph has an X-ray extent close to \emph{Chandra}'s 
PSF and that of M31N 2008-12a is only detected kinematically, DQ\,Her 
presents the best case for the detection of a resolved X-ray jet in a 
nova shell.  
We remark that, unlike RS\,Oph and M31N 2008-12a, which are recurrent 
novae, the nova shell of DQ\,Her is associated with a CV.

\section{SUMMARY}

We have presented the analysis of archival \emph{Chandra} ACIS-S and 
\emph{XMM-Newton} EPIC observations of the CV DQ\,Her. 
Our analysis has shown the presence of diffuse emission with a bipolar, 
jet-like morphology that extends up to distances 16$''$  from the 
progenitor star along the minor axis of the nova shell, thus protruding 
away from the nova shell.

We have also shown that the \emph{XMM-Newton} soft band image traces 
emission both from the jet and from a hot bubble filling the nebula 
around DQ\,Her. 
The latter has been formed as a result of an adiabatically-shocked blast 
wave very similar to supernova explosions.  
%A far-UV \emph{GALEX} image of DQ\,Her confirms that the hot bubble 
%is confined within the optical nova shell. 

The spectra of the extended X-ray emission is notably different to that 
of DQ\,Her, exhibiting the presence of emission lines from the 
O\,{\sc vii} triplet at 0.58~keV, the Ne and Fe complex at 0.9~keV, and 
Mg\,{\sc xi} at 1.4~keV. 
The bipolar structure has a plasma temperature of $2\times10^{6}$~K 
with an X-ray luminosity in the 0.3--5.0~keV energy range of 
$L_\mathrm{X,diff}=(2.1\pm1.3)\times10^{29}$~erg~s$^{-1}$. 
Its electron density and estimated mass are $n_\mathrm{e}\approx2$~cm$^{-3}$ 
and $m_\mathrm{X}\approx 3\times10^{-6}$~M$_{\odot}$, respectively.

We propose that the bipolar structure detected with \emph{Chandra} and 
\emph{XMM-Newton} is a jet. 
Its non-thermal emission component strongly supports that it is a a magnetized jet, 
arising as the result of the {\it hoop stress} mechanism observed in other 
stellar systems. 
Under this scenario the jet would be continuously fed by material 
that falls into the accretion disk and is then ejected by the hoop 
stress.
The S-shape morphology of the jet could then be associated with 
the precession of the accreting disk at the core of DQ\,Her or 
with erratic jet wobbling.

%We proposed that the bipolar structure detected with \emph{Chandra} and 
%\emph{XMM-Newton} is a jet that resulted from the latest nova explosion 
%$\sim80$~yr ago. 
%The presence of a non-thermal component in the emission of this jet 
%suggests that it is a magnetized jet whose formation may arise as the 
%result of the {\it hoop stress} mechanisms described by \citet{Livio1997} 
%and observed in other stellar systems 
%\citep[such as the case of the protostellar object HH80;][]{CG2010}. Under this scenario the jet would be %continuously produced by material that fell into the accretiong disk and then ejected by the hoop strees.

The capabilities of the up-coming \emph{Athena} X-ray satellite 
will be able to resolve the morphological and spectral components 
in DQ\,Her and will help bringing light into the scenario proposed 
by the present work.

\section*{Acknowledgements}

We would like to thank the referee for prompt revision and useful comments 
that helped improve the presentation of our results. We also thank C.\,Carrasco-Gonz\'alez for analysing the available JVLA observations of DQ\,Her. 
JAT, MAG and GR-L are supported by the UNAM Direcci\'{o}n General de Asuntos del Personal Acad\'{e}mico (DGAPA) though the Programa de Apoyo a Proyectos de Investigaci\'{o}n e Innovaci\'{o}n Tecnol\'{o}gica (PAPIIT) projects IA100318 and IA100720. MAG acknowledges support
from grant PGC2018-102184-B-I00, co-funded with FEDER funds. GR-L acknowledge support from Consejo Nacional de Ciencia y Tecnolog\'{i}a (CONACyT) and Programa para el Desarrollo Profesional (PRODEP) Mexico. LS acknowledge support from UNAM DGAPA PAPIIT project IN101819. This work has made extensive use of the NASA’s
Astrophysics Data System. ES thanks CONACyT (Mexico) for a research studentship. This work was based on observations obtained with {\it XMM–Newton}, an ESA science mission with instruments and contributions directly funded by ESA Member States and NASA. The scientific results reported in this article are based on observations made by the {\it Chandra} X-ray Observatory and published previously in cited articles. 
%This research is based on observations made with the {\it GALEX} mission, obtained from the MAST data archive at
%the Space Telescope Science Institute.



%\bibliographystyle{mnras}
%\bibliography{} % if your bibtex file is called example.bib

%% Alternatively you could enter them by hand, like this:

\end{document}